# Smartphone sensors and video analysis: two *allies* in the Physics laboratory *battle field*

M Monteiro[1,2], C Cabeza[2], C Stari[3] and A C Marti[2]
[1]*Universidad ORT Uruguay, Cuareim 1451, Montevideo, Uruguay*
[2]*Facultad de Ciencias, UdelaR, Iguá 4225, Montevideo, Uruguay*
[3]*Facultad de Ingeniería, UdelaR, Av. Julio Herrera y Reissig 565, Montevideo, Uruguay*

**Abstract.** Recently, two technologies: video analysis and mobile device sensors have considerable impacted Physics teaching. However, in general, these techniques are usually used independently. Here, we focus on a less-explored feature: the possibility of using supplementary video analysis and smartphone (or other mobile devices) sensors. First, we review some experiments reported in the literature using both tools. Next, we present an experiment specially suited to compare both resources and discuss in detail some typical results. We found that, as a rule, video analysis provides distances or angular variables, while sensors supplies velocity or acceleration (either linear or angular). The numerical differentiation of higher derivatives, *i.e.* acceleration, usually implies noisier results while the opposite process (the numerical integration of a temporal evolution) gives rise to the accumulation of errors. In a classroom situation, the comparison between these two techniques offers an opportunity to discuss not only concepts related to the specific experiment but also with the experimental and numerical aspects including their *pros* and *cons*.

## 1. Video analysis and smartphone sensors in the laboratory

In the last years, two technologies stormed in classrooms and Physics laboratories. One consists in making use of video analysis, *i.e.*, the –manual or automatic– tracking of objects, frame by frame, in digital videos, either obtained by the students themselves or downloaded from an open library or repository [1]. The other one, more recent, is based on modern mobile devices. Indeed, numerous smartphone-based physics experiments have been proposed in the literature (see for example recent issues of the Physics Education, The European Journal of Physics or The Physics Teacher journals). These experiments take advantage of the built-in smartphone sensors as the accelerometer, gyroscope (angular velocity sensor), magnetometer, proximeter, luxometer (ambient light sensor), pressure sensor (barometer), microphone among others. As a general rule, in the Physics-teaching literature an endless number of experiments is proposed based on one or the other technology but seldom using *simultaneously* both of them. As we consider that this is a window of opportunity not been fully exploited, here we discuss some experiments involving both video analysis and smartphone sensors and analyse future perspectives.

Smartphone usage has strongly expanded over recent years. Remarkably, their use goes significantly beyond the original purpose of talking on the phone. Indeed, it is everyday more frequent to use smartphones as clocks, cameras, agendas, music players or GPS. More remarkable is the habit, especially among young people, of bringing their smartphones every time and everywhere. From a physicist's point of view, it is impressive that smartphones usually incorporate several sensors, including accelerometer, angular velocity sensor, magnetometer, proximity or pressure sensor. Although these sensors are not supplied with educational intentions in mind, they can be employed in a wide range of physical experiments, especially in high school or undergraduate laboratories [2-5]. Moreover, experiments with smartphones can be easily performed in non-traditional places as playgrounds, gyms, travel facilities, among many others. All the possibilities that smartphones exhibit, foster students' interest in exploring, measuring and discovering the physical world around them.

Video analysis is an useful tool in Physics laboratories at many levels. There are several software packages available but one of them, Tracker [1], free and open, has become a standard tool in the framework of Physics teaching. In all the cases, the underlying idea is the possibility of analysing digital records of experiments, frame-by-frame, to obtain physical magnitudes. The most common example is the obtention of linear or angular coordinates, however, as shown in several papers this procedure can be extended to center-of-mass or normal coordinates in oscillatory problems and also as a modelling tool [6,7]. In this approach, Tracker numerically solves the differential equations of motion proposed by the user using appropriate numerical algorithms.

In the next Section we will briefly review, some experiments that make simultaneous use of both technologies. After that, in Sections 3 an experiment with a physical pendulum is presented and in Section 4 the results and specific issues, *pros* and *cons*, are discussed. Finally, in Section 5 we present the final remarks.

## 2. Experiments involving video analysis and sensors

In the scientific literature a few articles, using both video analysis and smartphone sensors, have been recently published. In one of the pioneering references [3], Castro-Palacio *et al* proposed the use of the acceleration sensor of a smartphone for the study of the uniform and uniformly accelerated circular motions and corroborated their results with the measures obtained by video recordings of the experiments and physical models.

Another worth mentioning reference by three of us [4] is focused on the physical (also known as compound) pendulum. This system was studied by means of the acceleration and rotation (gyroscope) sensors. The pendulum can rotate, giving full turns in one direction, or oscillate about the equilibrium position (performing either small or large oscillations). Radial and tangential acceleration and the angular velocity obtained with smartphone sensors allow to analyse the dynamics and, remarkable, taking profit of the simultaneous use of the acceleration and gyroscope sensors, trajectories in the phase space are directly obtained. The coherence of the measures obtained with the different sensors is corroborated with video analysis. This approach was extended in Ref. [6] to include the obtention of the (real) acceleration instead of an apparent acceleration (as it is obtained directly by the acceleration sensor). This, in fact, is a consequence of the equivalence principle which sates that a sensor fixed in a non-inertial reference frame cannot discern between a gravitational field and an accelerated system. In this case, we shown how to process acceleration values read by these sensors to substrate the gravitational component.

Besides classical mechanics, other proposed experiments involve electromagnetism or modern Physics. Recently, Pirbhai [8] measured the *e/m* (electron charge-to-mass) ratio using smartphones to measure the magnetic field strengths and phone cameras (and video analysis) to determine *e/m* with considerable accuracy. Simo [9] also proposed an experiment involving electromagnetism and, in this case, to verify Faraday-Lenz law. Dropping a piece of magnet through a methacrylate tube that crosses a coil, the measured electromagnetic field is related to the position of the magnet depending on time. Another example is Ref. [10], where the dynamics of a traditional toy, the yoyo, is investigated theoretically and experimentally using smartphone' sensors. In particular, the angular velocity is measured using the gyroscope. The experimental results are complemented thanks to a digital video analysis. The concordance between theoretical and experimental results is also discussed.

It is also worth mentioning, the experiment about the shape of a liquid surface in a rotating frame depending on the angular velocity [11]. This experiment consists on a fluid in a narrow rectangular container placed on a rotating table. A smartphone fixed to the rotating frame simultaneously records the fluid surface with the camera and also, thanks to the smartphone sensor, the angular velocity of the non-inertial frame of reference. The video analysis is used to obtain the surface's shape: concavity of the parabola and height of the vertex. Finally, let us mention a curious experiment about the discharge of a RC circuit [12]. It is shown there how to analyse using video analysis the discharge of the capacitor and correlate data with the information provided by sensors.

## 3. The physical pendulum experiment

One typical example of a physical system of paramount importance in high-school or undergraduate level is the compound pendulum (also known as physical pendulum) which consists of a rigid body that can freely rotate around a horizontal axis through a fixed center of suspension. This experiment is appropriate to be analysed using both video analysis and smartphone sensors. We consider a physical pendulum composed of a bicycle wheel with its axis fixed in a horizontal position around which the wheel rotates in a vertical plane, and a smartphone affixed to the outer edge of the tire, as shown in Figure 1.

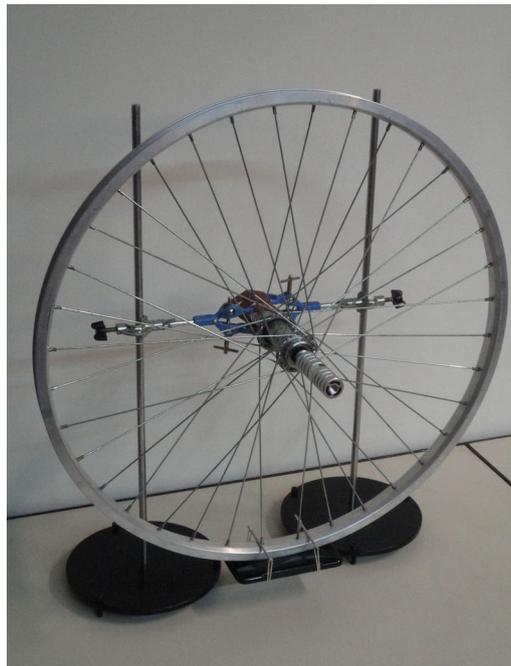

Figure 1. Experimental setup consisting on a bike wheel with the hub fixed on a horizontal axis and a smartphone fixed on the periphery of the tire.

An Android operated smartphone (LG G2 D805) furnished with a 3-axis LGE accelerometer sensor (STMicroelectronics, 0.001 m/s2 precision) and a 3-axis LGE gyroscope (STMicroelectronics, 0.001 rad/s precision) was used. Technical information regarding the exact location of the sensors within the smartphone was obtained from the manufacturer and verified by physical methods [5]. To communicate with the hardware (sensors), an appropriate piece of software, or *app*, is needed. In this experiment app *Androsensor* was used to record sensor readings [13]. Although, the sensor register the three components of the acceleration and the three of the angular velocity, only the radial, the tangential acceleration, and the angular velocity in the perpendicular plane are relevant in this experiment. According to the orientation of the smartphone shown in Fig. 1, these components are, respectively, $a_z$, $a_y$ and $\omega_x$. It is also necessary to setup the duration of the experiment and the sampling period. The registered data can be examined directly on the smartphone screen or transferred to the internet and studied using appropriate graphics packages.

As mentioned above, in addition to the smartphone sensors, video analysis was also used. With this objective, the experiment was recorded with a digital camera and the results examined using the already mentioned video analysis software Tracker [1] easily available in the internet.

### 4. Results and discussion
All the relevant physical variables, angle, angular velocity and angular acceleration, were obtained experimentally and compared using video analysis and the smartphone orientation sensor. In addition, the angle is calculated using the information provided by the accelerometer and the gyroscope as reported in [5]. A tracker screen-shot is displayed in Figure 2. As mentioned before, given length and time scales, the software analyses frame-by-frame and calculate, first of all, the angle. Next, using numerical schemes the angular velocity and, last, the angular acceleration.

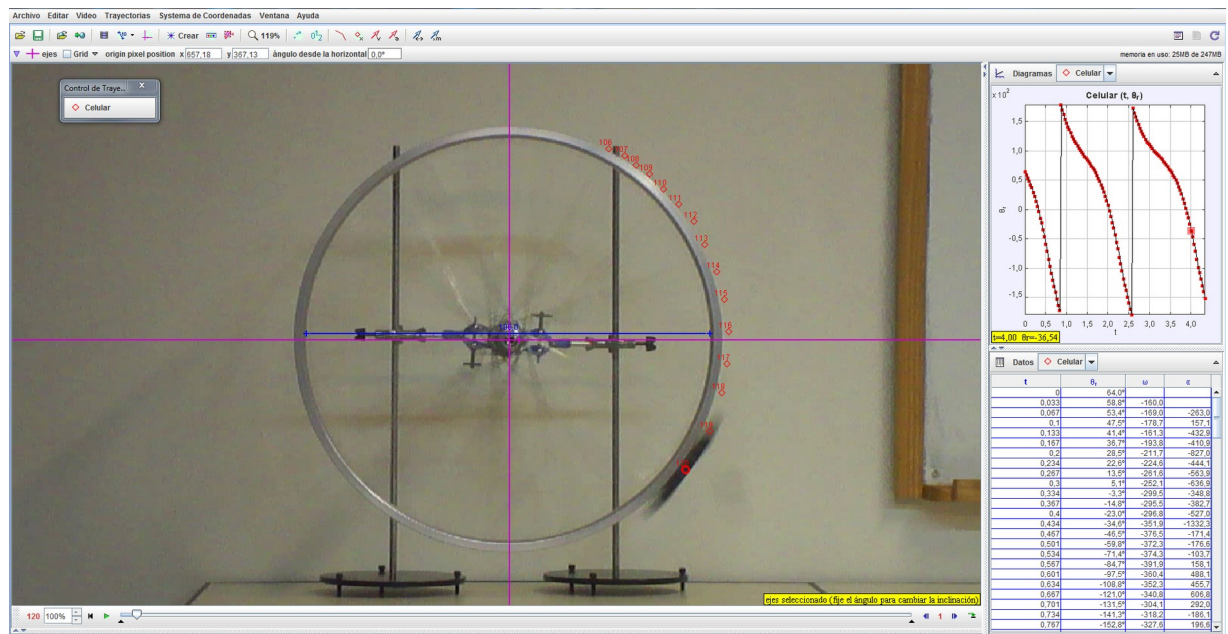

Figure 2. Tracker screen-shot. On the left the experimental system and the traces of a labelled point of the rim can be appreciated. On the right windows a graphical representation of the angle and the numerical values are also displayed.

The temporal evolution of the angle is depicted in Figs. 3 and 4. We distinguish two regions, in the first one (Figure 3) the pendulum passes through the uppermost position (the non-stable equilibrium point). Both the calculated coordinated and the video analysis agree very well, however, the orientation sensor is not able to provide correct values.

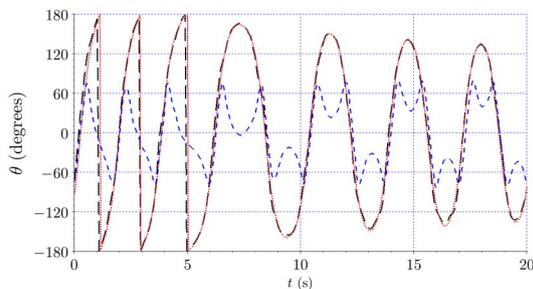

Figure 3. Angular variable as a function of time: black corresponds to calculated angles, red to video analysis and blue to orientation sensor. When the smartphone passes through the non-stable equilibrium point (uppermost position) the orientation sensor provides erroneous values.

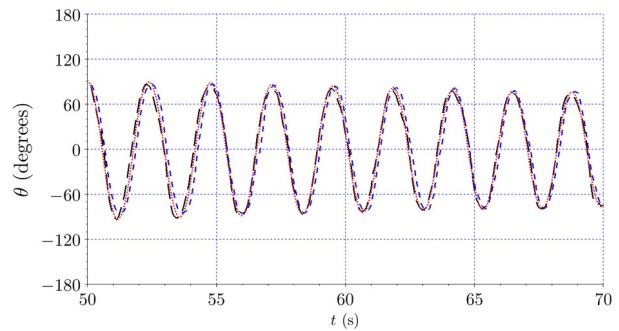

Figure 4. Angular variable as a function of time. Colour code as in Fig. 3. When the smartphone performs oscillations with amplitude below 90º does not pass through the uppermost position the three values agree very well.

Figure 5 shows the temporal evolution of the angular acceleration. As it can be appreciated the values obtained with video analysis (differentiating the angle twice) are considerably noisier than those obtained by the gyroscope (differentiating the angular velocity).

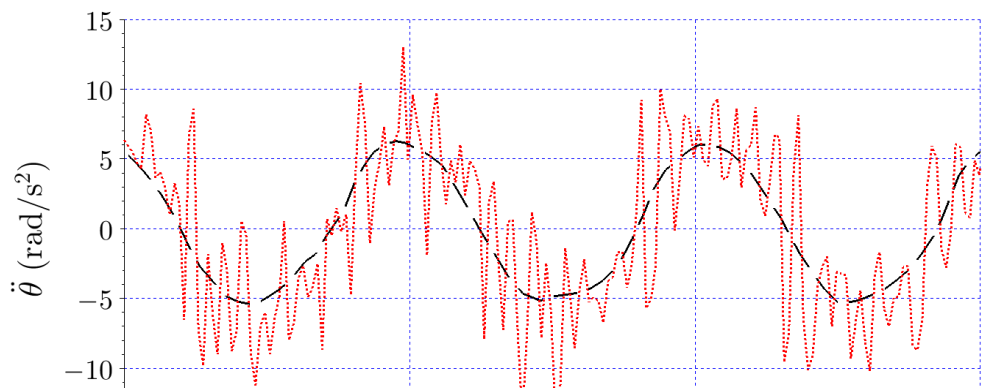

Figure 5. Angular acceleration as a function of time. The black dashed line corresponds to gyroscope measurements while the red points to the results of video analysis. The initial time is arbitrary.

## 5. Final remarks

In this article we have shown that smartphone sensors and video analysis are complementary tools in Physics-teaching. Both of them present their advantages and disadvantages and they can be employed in a supplementary way. As shown in several examples, available in the literature, these tools can be used, in specific experimental setups, to corroborate the coherence of different measures. In other experimental situations these tools can be used to obtain different physical quantities. In these cases, it is necessary to proceed with caution. Usually, physical quantities obtained with numerical schemes, as velocity or acceleration, obtained from video analysis are considerably noisier than those obtained with sensors. In addition, measures obtained with the linear acceleration sensor were found to be inaccurate, in particular when the smartphone moves in proximity to the point of stable equilibrium.

Both video analysis and smartphone sensors are valuable tools useful to elucidate a wide range of physical phenomena. An adequate understanding of the underlying operation principles is of fundamental importance, a fact which gains in significance as the use of smartphones and video cameras becomes more widespread with the expected decrease in cost. All the possibilities that these new technologies exhibit, foster students' interest in exploring, measuring and discovering the physical world around them.

**Acknowledgements.** We acknowledge P*rograma de Desarrollo de las Ciencias Básicas (PEDECIBA)* and *CSIC, UdelaR, Grupos I+D, Física Nolineal* (Uruguay) for financial support.